\documentclass[a4paper,showpacs,nofootinbib,floatfix,superscriptaddress, 12pt]{revtex4}
\usepackage{graphicx}
\usepackage{amsfonts}
\usepackage{amsmath}
\usepackage{hyperref}
\usepackage{float}

\renewcommand{\vec}[1]{\mathbf{#1}}
\def\be{\begin{equation}}
\def\ee{\end{equation}}
\def\bea{\begin{eqnarray}}
\def\eea{\end{eqnarray}}

\begin{document}

\title{Extraction of shear viscosity in stationary
states of relativistic particle systems}

\author{F.\ Reining}
\author{I.\ Bouras}
\author{A.\ El}
\author{C.\ Wesp}
\affiliation{Institut f\"ur Theoretische Physik,
Johann Wolfgang Goethe-Universit\"at,
Max-von-Laue-Str.\ 1, D-60438 Frankfurt am Main, Germany}

\author{Z.\ Xu}
\affiliation{Institut f\"ur Theoretische Physik,
Johann Wolfgang Goethe-Universit\"at,
Max-von-Laue-Str.\ 1, D-60438 Frankfurt am Main, Germany}
\affiliation{Frankfurt Institute for Advanced Studies,
Ruth-Moufang-Str.\ 1, D-60438 Frankfurt am Main, Germany}

\author{C.\ Greiner}
\affiliation{Institut f\"ur Theoretische Physik,
Johann Wolfgang Goethe-Universit\"at,
Max-von-Laue-Str.\ 1, D-60438 Frankfurt am Main, Germany}

\begin{abstract}

Starting from a classical picture of shear viscosity we
construct a stationary velocity gradient in a microscopic
parton cascade. Employing the Navier-Stokes ansatz we
extract the shear viscosity coefficient $\eta$. For elastic
isotropic scatterings we find an excellent agreement
with the analytic values. This confirms the applicability
of this method. Furthermore for both elastic and inelastic
scatterings with pQCD based cross sections we extract the shear
viscosity coefficient $\eta$ for a pure gluonic system and find 
a good agreement with already published calculations.

\end{abstract}


\date{\today}

\maketitle

\section{Introduction}
\label{sec:intro}

Recent results of the Relativistic Heavy Ion Collider (RHIC) 
and of the Large Hadron Collider (LHC)
indicate the formation of a new state of matter, 
the quark-gluon plasma (QGP), in relativistic
heavy-ion collisions. The large value of the elliptic flow 
coefficient $v_2$ observed in these experiments \cite{star,phen,phobos,Aamodt:2010pa}
leads to the indication that the QGP behaves like 
a nearly perfect fluid.
This has been confirmed by calculations of viscous
hydrodynamics \cite{hs08, Luzum:2008cw, Heinz:2009xj, Teaney:2009qa, Niemi:2011ix, Schenke:2011tv, Song:2011qa} and microscopic
transport calculations \cite{Xu:2007jv, Ferini:2008he}.
However, the shear viscosity coefficient $\eta$ 
has a finite value, possibly close to the conjectured lower bound 
$\eta / s = 1 / 4 \pi$ from the correspondence between
conformal field theory and string theory in an Anti-de-Sitter space \cite{Kovtun:2004de}.
In comparison to ideal hydrodynamic calculations \cite{Kolb:2000fha}, 
dissipative hydrodynamic formalisms with finite $\eta/s$ ratio \cite{hs08, Luzum:2008cw, Heinz:2009xj, Teaney:2009qa, Niemi:2011ix, Schenke:2011tv, Song:2011qa} demonstrate a better 
agreement of the differential elliptic flow $v_2(p_t)$ with experimental data. The shear viscosity
is therefore an important parameter in viscous hydrodynamics but needs to be calculated from 
microscopic theory.

The $\eta/s$ ratio was obtained in a full leading order pertubative QCD
calculation in Ref.~\cite{AYM}. The Boltzmann-Vlasov equation and quasi-particle picture 
were recently employed to calculate the $\eta/s$ ratio of a gluon gas in Ref.~\cite{Bluhm:2010qf}.
The shear viscosity coefficient has also been extracted from
microscopic transport calculations with
BAMPS (Boltzmann Approach of Multi Parton Scatterings) simulations \cite{Xu:2004mz, Xu:2007aa}
using expressions based on a first-order gradient expansion of the
Boltzmann Equation \cite{ Xu:2007ns} and the entropy principle underlying the
second-order Israel-Stewart hydrodynamics \cite{El:2008yy}.

\begin{figure}[h]
\includegraphics[width=8.5cm]{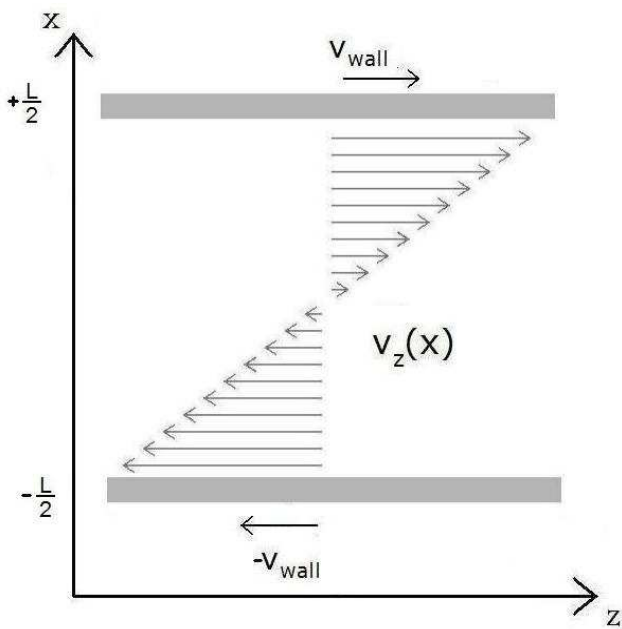}
\caption{The classical definition of shear viscosity. Two plates 
	moving in opposite directions with velocity
	$\pm v_{\rm wall}$. A flow gradient is established
	between the plates. The viscosity is proportional
	to the frictional force.}
\label{fig:geometry}
\end{figure}

The goal of this work is to extract the shear viscosity
coefficient $\eta$ numerically from microscopic calculations 
using a standard setup motivated by 
the classical textbook picture \cite{reif, Reining_diploma}.
In Fig.~\ref{fig:geometry} we introduce a 
particle system embedded between two plates. 
The two plates move in
opposite directions each with velocity $v_{\rm wall}$ in
$z$-direction. The moving walls are supplemented
by two thermal reservoirs with $\pm v_{\rm wall}$.
In $x$-direction the system has an extension of size $L$.
In $y$- and $z$-direction the system
is homogeneous and can be of infinite size.
The mean free path of the particles should be very
small compared to the system size,
i.e. 
$\lambda_{\rm mfp} < < L$. On a sufficiently
long time-scale a stationary velocity field
$v_{\rm z}(x)$ should be established.
In the non-relativistic limit the velocity
field is linear. With the Navier-Stokes-ansatz
the shear stress tensor $\pi^{\mu \nu}$ is proportional
to the gradient of the velocity
\begin{equation}
 \pi^{xz}= - \eta \frac{\partial v_{\rm z}(x)}{\partial x} \, .
\label{eq:eta}
\end{equation}
The proportionality factor is defined to be the
shear viscosity coefficient $\eta$.
In Sec.~\ref{sec:definitions} we give basic definitions and information on the 
numerical model we use.
In Sec.~\ref{sec:grad} we demonstrate that
Eq.~\eqref{eq:eta} does not hold in general for the relativistic
case, where the gradient is not necessarily linear and 
we discuss the shape of an ideal relativistic velocity gradient.
Furthermore we will discuss the effect of viscosity
and finite size effects on the velocity profile in
Sec.~\ref{sec:grad}, where an analytical formulation for the
shape of the velocity profile is derived.
We employ BAMPS to reproduce the velocity gradient
as discussed in this chapter.
In Sec.~\ref{sec:viscosity} we compare the numerical results
for the shear viscosity coefficient $\eta$ to an analytical value in order to confirm
the applicability of our method.
Finally we present the results on 
shear viscosity to entropy density ratio obtained from BAMPS with 
cross sections based on pertubative quantum chromodynamics (pQCD)
and compare them to existing calculations.
We close with a summary.

\section{Basic Idea and Definitions}
\label{sec:definitions}

When systems are in stationary states, the first-order Navier-Stokes 
formulation of relativistic viscous hydrodynamics 
can be used to calculate the shear viscosity $\eta$, which is the
proportionality factor between the shear tensor
$\pi^{\mu\nu} = T^{ \langle \mu\nu \rangle}$ and
the velocity gradient $\nabla^{\langle \mu} u^{\nu \rangle}$:
\begin{equation}
\label{NS}
 \pi^{\mu\nu} = 2 \eta \nabla^{<\mu} u^{\nu>} \, ,
\end{equation}
where the projection
\begin{equation}
B^{\langle \mu \nu \rangle} 
\equiv \left[\frac{1}{2}\left( \Delta^{\mu}_{\alpha} \Delta^{\nu}_{\beta} +
\Delta^{\nu}_{\alpha} \Delta^{\mu}_{\beta} \right) - \frac{1}{3}
\Delta^{\mu \nu} \Delta_{\alpha \beta} \right] B^{\alpha \beta}
\end{equation}
denotes the symmetric traceless part of the tensor
$B^{\mu \nu}$. $\Delta^{\mu \nu} = g^{\mu\nu} - u^\mu  u^\nu$
is the transverse projection operator and the metric is
$g^{\mu \nu} = \textrm{diag}(1,-1,-1,-1)$.

Some definitions are in order. We use the Landau's definition
of the hydrodynamic four-velocity \cite{Landau_book}:
\begin{equation}
u^\mu = \frac{T^{\mu \nu} u_\nu}{e} \,,
\end{equation}
where
\begin{equation}
\label{kinetic:T_mu_nu}
T^{\mu \nu} = \int \frac{d^3p}{(2\pi)^3 p^0} \, p^{\mu} p^{\nu} f(x,p)
\end{equation}
is the energy-momentum tensor and the local energy density is defined as
\begin{equation}
e = u_{\mu} T^{\mu \nu} u_{\nu} \,.
\end{equation}
The shear tensor $\pi^{\mu\nu}$ is the difference of $T^{\mu\nu}$
to its equilibrium value. For the geometry depicted in Fig.~\ref{fig:geometry}
$u^\mu=\gamma(1,0,0,v_z)$ with $\gamma=1/\sqrt{1-v_z^2}$.

We will build up stationary states of particle systems
via numerical simulations, which are realized
by employing the microscopic transport model BAMPS, which solves the Boltzmann equations
for on-shell particles within a stochastic model \cite{Xu:2004mz,Xu:2007aa}.
In principle any microscopic transport model can be used for this purpose.

Local values of $\pi^{\mu\nu}$ and $u^\mu$ can be easily extracted 
from the numerical simulations by averaging over all particles contained in
a bin of size $\Delta x$. However, to obtain the gradient of $u^\mu$
one has to take values from neighbouring local cells, which would
cause additional numerical errors. To avoid such numerical problem
we will first derive the analytical form of 
$v_z(x)$ for the given setup in Fig.~\ref{fig:geometry}. Then we
use this form and the numerically extracted $\pi^{\mu\nu}$ to calculate
the shear viscosity.

\section{Velocity, rapidity and finite size effect}
\label{sec:grad}
\subsection{Analytical Derivation}


Instead of the hydrodynamic velocity $v_z(x)$ we address the position
dependence of the rapidity $y(x)$, which is defined by
\begin{equation}
\label{rapidity}
y(x)=\frac{1}{2} \ln \frac{1+v_z(x)}{1-v_z(x)}\,.
\end{equation}
Thus, $v_z(x)=\tanh y(x)$. In the non-relativistic limit, where $v_z(x)$ is
small, we have $v_z(x)\approx y(x)$.
The advantage of $y(x)$ is that it gets a shift by a Lorentz-boost
e.g. with $v_z(x_A)$
\begin{equation}
\Lambda_{v_z(x_A)} [y(x)]=y(x)-y(x_A)\,.
\end{equation}
Demanding boost-invariance, i.e., $\Lambda_{v_z(x_A)}[y(x)]=y(x-x_A)$,
we obtain the solution $y(x)=ax+b$, where $a$ und $b$ are constant. Due to
the boundary condition $y(x=\pm L/2)=\pm y_{wall}$, $y(x)$ is symmetric in x
and thus, $b=0$. If $y(x)$ is continuous at the boundaries, we have
\begin{equation}
\label{ideal}
y(x)=\frac{2y_{wall}}{L}\ x \,.
\end{equation}

In the following we will convince ourselves from relativistic kinetic theory
that Eq.~\eqref{ideal} is only valid if the particle mean free path vanishes,
or the distance $L$ between two plates is infinitely long. For a non-vanishing
mean free path and a finite distance $L$ we will see discontinuities of
$y(x)$ at the boundaries. This is referred to as a finite size effect.

We consider a general local observable $A(x,t)$ with the definition
\begin{equation}
\label{observ}
A(x,t)=\frac{1}{n(x,t)}\int d\Gamma_1 F_A(p_1) f(p_1;x,t) \,,
\end{equation}
where $d\Gamma_1=d^3p_1/(2\pi)^3$ and $n(x,t)=\int d\Gamma_1 f(p_1;x,t)$
is the particle number density. $p$ denotes the particle four-momentum.
In our case $n$ does not depend on position and time.
In particular, for $F_A(p_1)=n p^\mu_1/p^0_1$ we have the definition 
of particle four-flow $A(x,t)=N^\mu(x,t)$; 
for $F_A(p_1)=\frac{1}{2}\ln [(p^0_1+p^z_1)/(p^0_1-p^z_1)]$
we obtain the rapidity $A(x,t)=y(x,t)$ as given in Eq.~\eqref{rapidity},
when using the Landau definition of the hydrodynamic four-velocity.
For stationary states $A(x,t)$ and the particle distribution function
$f(p;x,t)$ are constant in time.

We define $\tilde f(p;x,t)=f(p;x,t)/n(x,t)$, which is the probability
density for the occurrence of a particle with momentum $p$ around $d\Gamma$
at $(x,t)$. One obtains $\tilde f(p;x,t)$ by summing probabilities for such
events that a collision at $(x',t')$ makes a particle having the
momentum $p$ and this particle travels to $x$ at $t$ without further
collisions. It is mathematically expressed by
\begin{eqnarray}
\label{probab1}
\tilde f(p_1;x,t)&=&\theta(p_{1x}) \int_{-\infty}^x dx' 
w_{gain}(p_1;x',t') w_{free}(p_1;x',t';x,t) + \nonumber \\
&&\theta(-p_{1x}) \int_x^{\infty}
dx' w_{gain}(p_1;x',t') w_{free}(p_1;x',t';x,t) \,,
\end{eqnarray}
where $w_{gain}(p_1;x',t')$ denotes the probability density that a particle
with momentum $p_1$ is created via a collision at $(x',t')$, and
$w_{free}(p_1;x',t';x,t)$ the probability that this particle travels
from $(x',t')$ to $(x,t)$ without further collisions.
Because $\tilde f(p;x,t)$ is invariant under the transformation
$p\to -p$, the two integrals in Eq.~\eqref{probab1} are equal. Thus,
\begin{equation}
\label{probab2}
\tilde f(p_1;x,t)=\frac{1}{2} \int_{-\infty}^{\infty} dx' 
w_{gain}(p_1;x',t') w_{free}(p_1;x',t';x,t) \,.
\end{equation}
Our goal is to find the relation between $A(x,t)$ and $A(x',t')$,
which then can be used to solve $A(x,t)$ analytically when the boundary
conditions are given.

Using the standard definition of cross section for binary collisions
of identical particles
\begin{equation}
\sigma_{22}=\frac{1}{4s}\int \frac{d\Gamma_1}{2p^0_1}\frac{d\Gamma_2}{2p^0_2}
|M_{1'2'\to 12}|^2 (2\pi)^4 \delta^{(4)}(p'_1+p'_2-p_1-p_2) \,,
\end{equation}
where $M_{1'2'\to 12}$ is the matrix element and $s=(p_1+p_2)^2=(p'_1+p'_2)^2$
is the invariant mass, we have
\begin{equation}
\label{wgain}
w_{gain}(p_1;x',t') dx'=\frac{1}{n}\int d\Gamma'_1 d\Gamma'_2
f(p'_1;x',t')f(p'_2;x',t') v_{rel} \frac{d\sigma_{22}}{d\Gamma_1} dt' \,.
\end{equation}
$v_{rel}=s/(2p'^0_1p'^0_2)$ denotes the relative velocity for massless
particles. $dt'$ is the average time interval, during which a particle
travels through $dx'$: $dt'=dx' <|p'_x|/p'_0>^{-1}$ and $<|p'_x|/p'_0>=1/2$
in thermal equilibrium.

The probability $w_{free}(p_1;x',t';x,t)$ is a product of 
$w_{free}(p_1;x'',t'';x''+dx'',t''+dt'')$ over $x''$ from $x'$ to $x$:
\begin{equation}
\label{wfree1}
w_{free}(p_1;x',t';x,t)=\prod_{x''=x'}^x 
w_{free}(p_1;x'',t'';x''+dx'',t''+dt'')
=\prod_{x''=x'}^x [1-w_{loss}(p_1;x'',t'')dx'']\,.
\end{equation}
$w_{loss}(p_1;x'',t'')$ denotes the probability density that a particle
with momentum $p_1$ is destroyed via a collision at $(x'',t'')$
and is expressed by
\begin{equation}
\label{wloss1}
w_{loss}(p_1;x'',t'')dx''=
\int d\Gamma_2 f(p_2;x'',t'') v_{rel} \sigma_{22} dt'' \,,
\end{equation}
where $v_{rel}=s/(2p^0_1p^0_2)$ and $dt''=dx'' (p_1^x/p_1^0)^{-1}$.

We now approximate $w_{loss}(p_1;x'',t'')$ to be the averaged one over $p_1$:
\begin{equation}
\label{wloss2}
w_{loss}(p_1;x'',t'')dx'' \approx
\int d\Gamma_2 f(p_2;x'',t'') \langle v_{rel} \sigma_{22} \rangle
\langle \frac{|p_1^x|}{p_1^0} \rangle^{-1} dx''=
2n\langle v_{rel} \sigma_{22} \rangle dx''=\frac{2dx''}{\lambda_{mfp}}\,,
\end{equation}
where $\lambda_{mfp}$ denotes the particle mean free path.
This approximation applies for isotropic cross sections. In general,
if the angular distribution is non-isotropic 
$\lambda_{mfp}$ has to be replaced by an effective length scale, 
which is calculated as an average of the differential cross section.
With Eq.~\eqref{wloss2} we obtain the obvious expression
\begin{equation}
\label{wfree2}
w_{free}(p_1;x',t';x,t)=\lim_{dx''\to 0} \left ( 
1- \frac{2dx''}{\lambda_{mfp}} \right )^{|x-x'|/dx''}=
\exp\left ( -\frac{2|x-x'|}{\lambda_{mfp}} \right )\,.
\end{equation}

Putting Eqs.~\eqref{probab2}, \eqref{wgain}, and \eqref{wfree2} into
Eq.~\eqref{observ} gives
\begin{equation}
A(x,t)=\int_{-\infty}^{\infty} dx' e^{-\frac{2|x-x'|}{\lambda_{mfp}}}
\frac{1}{n}\int d\Gamma'_1 d\Gamma'_2 f(p'_1;x',t')f(p'_2;x',t')
v_{rel} \int d\Gamma_1 F_A(p_1) \frac{d\sigma_{22}}{d\Gamma_1} \,.
\end{equation}
It is clear that replacing $F_A(p_1)$ by $F_A(p_1)+F_A(p_2)$ will
leads to $2A(x,t)$. We now consider particular observables $A(x,t)$
such that $F_A$ is conserved in each collision, i.e.,
$F_A(p'_1)+F_A(p'_2)=F_A(p_1)+F_A(p_2)$. We then have
\begin{eqnarray}
\label{main}
A(x,t)&=&\int_{-\infty}^{\infty} dx' e^{-\frac{2|x-x'|}{\lambda_{mfp}}}
\frac{1}{n}\int d\Gamma'_1 d\Gamma'_2 f(p'_1;x',t')f(p'_2;x',t')
F_A(p'_1) v_{rel} \int d\Gamma_1 \frac{d\sigma_{22}}{d\Gamma_1} \nonumber\\
&\approx&\int_{-\infty}^{\infty} dx' e^{-\frac{2|x-x'|}{\lambda_{mfp}}}
\frac{1}{n}\int d\Gamma'_1 F_A(p'_1)f(p'_1;x',t')
\int d\Gamma'_2 f(p'_2;x',t') \langle v_{rel} \sigma_{22} \rangle \nonumber\\
&=&\frac{1}{\lambda_{mfp}} 
\int_{-\infty}^{\infty} dx' e^{-\frac{2|x-x'|}{\lambda_{mfp}}} A(x',t') \,.
\end{eqnarray}
The same approximation is made as for $w_{loss}$ in Eq.~\eqref{wloss2}.
Equation~\eqref{main} resembles the one derived in Ref.~\cite{reif}
using "path integral method" in non-relativistic cases.

We emphasize that Eq.~\eqref{main} holds only if the total $F_A$ is conserved
in collisions. For instance, the total particle velocity 
$\vec{p_1}/E_1+\vec{p_2}/E_2$ is not conserved except in case the energy of all
particles is same, whereas the total particle momentum rapidity is conserved.
Therefore, the rapidity $y(x)$ defined by Eq.~\eqref{rapidity} obeys 
Eq.~\eqref{main}, but the hydrodynamic velovity $v_z(x)$ does not.
However, the total particle momentum rapidity is not conserved in
$2\to 3$ or $3\to 2$ processes. In this case one has to take detailed
balance into account and the sum of the total rapidity of a $2\to 3$ and
its back reaction is conserved on average. If $y(x)$ is conserved in 
collisional processes, it obeys Eq.~\eqref{main}.

Equation~\eqref{main} represents a homogeneous first-order integral 
equation for $A(x)$. It can easily be shown that the second derivative
of $A(x)$ vanishes, which leads to the solution $A(x)=ax+b$, where $a$ and
$b$ are constant.
We choose the boundary conditions to be $A(x) =  -y_{wall}$ for $x <- L/2$ and
$A(x) =  y_{wall}$ for $x > L/2$ to reproduce the scenario
indroduced in Sec.~\ref{sec:intro}. Since this scenario is symetric in $x$ we have $b=0$.
To determine $a$ we insert $A(x)= ax$ into Eq.~\eqref{main}
and obtain $a = 2y_{wall}/(L+\lambda_{mfp})$.
Finally the rapidity has the following form
\begin{equation}
\label{rap}
 y(x) = \frac{2y_{wall}}{L+\lambda_{mfp}}\ x \,.
\end{equation}
We recognize the discontinuities of $y(x)$ at the boundaries, which
disappear only for vanishing mean free path $\lambda_{mfp} \to 0$ or
long distance $L\to \infty$. 
Equation~\eqref{rap} is a new finding and accounts for finite size 
effects which must be taken into accout, if for numerical reasons
$\lambda_{mfp}/L$ cannot be made sufficiently small.

\subsection{Numerical Confirmation}
In this subsection we will confirm our finding Eq.~\eqref{rap}
by performing numerical transport calculations. We employ
the parton cascade BAMPS. Details of numerical operations
can be found in Refs.~\cite{Xu:2004mz,Xu:2007aa}.
One important feature of BAMPS is that the model can simulate
multiplication and annihilation processes such as 
the gluon bremsstrahlung process and its back reaction 
$gg \leftrightarrow ggg$ with full detailed balance.
In order to verify the analytic findings we will first employ
isotropic cross sections in BAMPS in the following.

The numerical realization of the boundary conditions is as follows.
Particles that reach the boundaries $x=\pm L/2$ are removed, which
simulates the particle absorption by the plates. Independent of
the absorption, the plates emit particles, which pick up the
velocities $\pm v_{wall}$ of the plates. Here we treat the plates as
thermal reservoirs of particles with the same temperature as
those between the plates. 
The momentum distribution for emitting particles
is proportional to the equilibrium Boltzmann distribution $f_{wall}(p)$
and the particle velocity $p_x/E$:
\begin{equation}
\frac{dN_{em}}{dtd^3p} \sim \frac{p_x}{E}f_{wall}(p)
\end{equation}
with
\begin{equation}
\label{boltzmann}
f_{wall}(p)=g\,e^{-\frac{p_\mu u_{wall}^\mu}{T}}\,,
\end{equation}
where $u_{wall}^\mu=\gamma_{wall}(1,0,0,v_{wall})$, 
$\gamma_{wall}=1/\sqrt{1-v_{wall}^2}$, $g=16$ is the degeneracy factor for gluons in $SU(3)$,
and $T$ is the temperature.
In the distribution~\eqref{boltzmann} we neglect the quantum 
statistic factor for bosons and fermions. 
The rate of emissions
can be calculated analytically (see App.~\ref{section_app:wall})
and is
\begin{equation}
\frac{dN_{em}}{dt} =  \frac{1}{4} A_{wall} \, n_{wall} \, ,
\end{equation}
where $A_{wall}$ is the transverse area of the plates and $n_{wall}$ is
the particle density. 
In the $xy$- and $xz$-plane the boundary conditions are periodic.

\begin{figure}[h!]
\includegraphics[width=8.5cm]{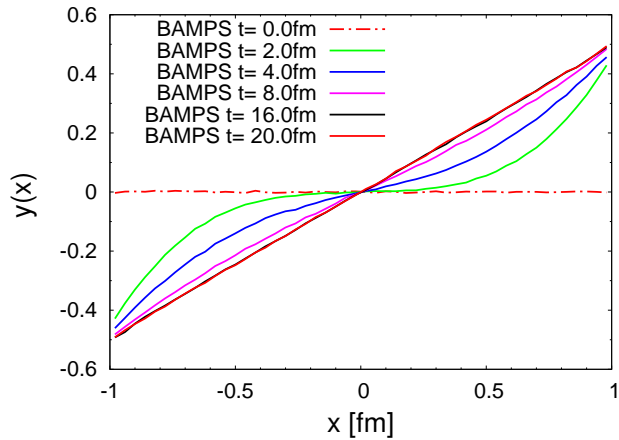}
\caption{Build-up with time of the rapidity profile. Results are obtained by
averaging $500$ events.
}
\label{fig:relax}
\end{figure}

Particles between the two plates are initially distributed by the
equilibrium form like Eq.~\eqref{boltzmann} with zero velocity.
Figure~\ref{fig:relax} shows the buildup of the rapidity profile.
We have chosen $L=2$ fm, $v_{wall} = 0.5$ ($y_{wall}=0.55$), 
and $T = 0.4$ GeV. Only binary collisions with a constant cross section
are considered. The mean free path is set to be $\lambda_{mfp}=0.2$ fm.
Collision angles of the binary scatterings are distributed isotropically.

The timescale for the buildup of the rapidity (or velocity) profile can be
estimated as the mean diffusion time of particles traveling from one plate 
to another. For a gaussian diffusion process one has \cite{reif}
\begin{equation}
<x^2>= 2 D t \,,
\end{equation}
where in the non-relativistic limit the diffusion constant $D$ is the
ratio of the shear viscosity $\eta$ to the mass density $\rho$.
For relativistic case we replace $\rho$ by the energy density $e$.
As we will see in the next section, 
$\eta \approx 1.2654 n T \lambda_{mfp}=0.42 e \lambda_{mfp}$
(see Eq.~\eqref{eq:etans}), where $e=3nT$ is used. Thus,
\begin{equation}
\label{relaxtime}
t =\frac{L^2 }{2D}=\frac{L^2 e}{2\eta}\approx \frac{L^2}{0.82\lambda_{mfp}}\,.
\end{equation}
For our setup we find $t\approx 24$ fm/c, which is consistent with
the numerical results shown in Fig.~\ref{fig:relax}.

\begin{figure}[h!]
\includegraphics[width=8.5cm]{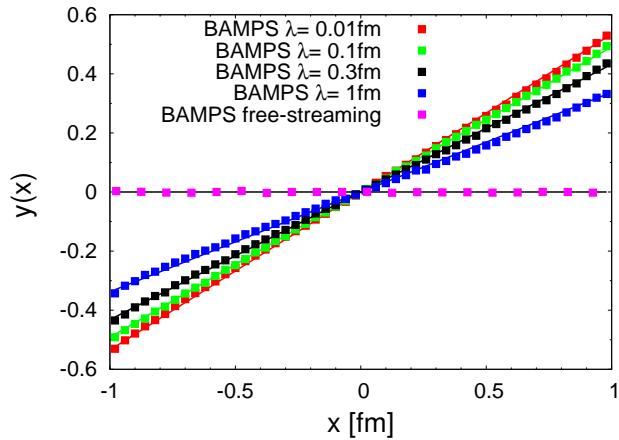}
\caption{Rapidity profiles for different mean free paths,
$\lambda_{\rm mfp} = 0.02, 0.2, 2, \infty ~\rm{ fm}$.
Constant cross sections and isotropic distribution of the
collision angle are considered. The numerical results from BAMPS
(symbols) are compared with the analytical ones (lines) 
given by Eq.~\eqref{rap}.
}
\label{fig:meanfreepath}
\end{figure}

Figure~\ref{fig:meanfreepath} shows the final rapidity profiles
at sufficient long times.
Calculations are performed for several mean free paths, in order to
demonstrate the finite size effect. The differential cross sections are 
momentum-independent, i.e. isotropic. 
The lines present the analytical results given via Eq.~\eqref{rap}, 
while the symbols show the numerical values.
One can see an excellent agreement, although approximations are made to
obtain Eq.~\eqref{rap}. This indicates the validity of the
approximations for using isotropic cross sections.

On the contrary, when using the pQCD cross sections for gluons,
which strongly depend on the invariant mass $s$ 
one will have deviations from Eq.~\eqref{rap}.
The elastic and gluon bremsstrahlung process and its back reaction implemented in BAMPS
 are based on pQCD matrix elements given in Ref.~\cite{Xu:2004mz}.
Although the numerically extracted rapidity profile is different from
the analytical form Eq.~\eqref{rap}, it is still linear in $x$.
Replacing $\lambda_{mfp}$ in Eq.~\eqref{rap} with
an effective scale $\lambda_{eff}$ one obtains the general formula.
Using pQCD cross sections $\lambda_{eff}$ has to be extracted numerically.
Qualitatively, $\lambda_{eff}$ for pQCD interactions should be
larger than $\lambda_{mfp}$, since pQCD-based processes prefer
small angle scatterings and thus, are not as efficient for momentum
transport as scatterings with isotropic angular distribution.

\section{Extraction of Shear Viscosity}
\label{sec:viscosity}

In stationary states we can use the Navier-Stockes's formula 
Eq.~\eqref{NS} to calculate the shear viscosity $\eta$.
For the particular setup shown in Fig.~\ref{fig:geometry},
Eq.~\eqref{NS} is simplified to
\begin{equation}
\pi^{xz}=-\eta \frac{d\gamma v_z(x)}{dx}
\end{equation}
with $\gamma=1/\sqrt{1-v_z^2(x)}$.
Using Eq.~\eqref{rap} for $v_z(x)=\tanh y(x)$ we obtain
\begin{equation}
\label{extract}
\eta=-\pi^{xz} \sqrt{1-v_z^2(x)}\, \frac{L+\lambda_{eff}}{2y_{wall}}\,.
\end{equation}
(Here $\lambda_{mfp}$ replaced by $\lambda_{eff}$.)
$\pi^{xz}$ and $v_z(x)$ are extracted from BAMPS (the results of
$v_z(x)$ are already shown in the previous section), whereas $\lambda_{eff}$
is obtained by fitting $y(x)$.

Results for isotropic constant cross sections are presented in Sec.~\ref{sec:iso}.
Section~\ref{sec:pQCD} contains results for full pQCD interactions.

\subsection{Elastic isotropic constant cross sections}
\label{sec:iso}

Elastic isotropic constant cross sections are meant that
cross sections for elastic binary collisions are constant
and the distribution of collision angle is isotropic.
In this case the shear viscosity of an ultrarelativistic Maxwell-Boltzmann gas 
is well known \cite{deGroot}:
\begin{equation}
 \eta^{NS} \approx 1.2654 \,\frac{T}{\sigma} =
1.2654 \,n T \,\lambda_{\rm mfp}\,.
\label{eq:etans}
\end{equation}
Equation~\eqref{eq:etans} serves as a benchmark to check the numerical
methods applied to calculate shear viscosity.

\begin{figure}[h]
\includegraphics[width=8.5cm]{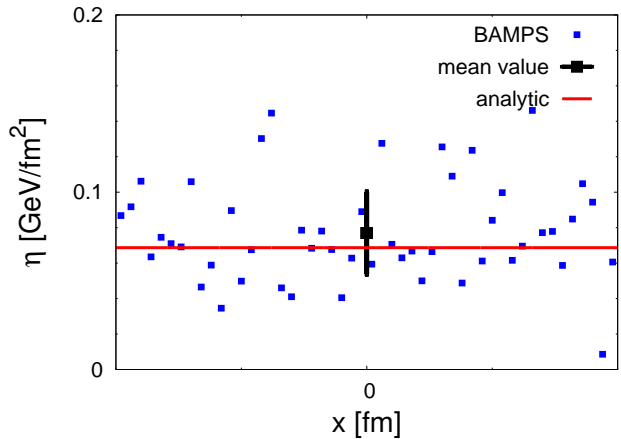}
\caption{Shear viscosity extracted from BAMPS
and compared to the analytical result
from Eq.~\eqref{eq:etans} for $\lambda_{\rm mfp} = 0.02$ fm.}
\label{fig:eta2}
\end{figure}

Setups for numerical calculations are $L=2$ fm, $v_{wall} = 0.5$,
and $T=0.4$ GeV. Results are averaged over $N_{events}=2000$ events.
Figure~\ref{fig:eta2} shows the numerically extracted
shear viscosity from Eq.~\eqref{extract} in each bin of 
size $\Delta x=0.2~\rm{fm}$.
The mean value is in good agreement with the analytical one from Eq.~\eqref{eq:etans}
 within the standard deviation, which decreases with $1/\sqrt{N_{events}}$.

Of couse this method for the extraction of shear viscosity can only be applied, when
the particle system has relaxed to a stationary state. The relaxation
time can be estimated according to Eq.~\eqref{relaxtime}. The relaxation time is
inversely proportional to the mean free path and thus the
extraction of shear viscosity for nearly perfect fluids with
high $N_{events}$ is time-consuming.

\begin{figure}[h]
\includegraphics[width=8.5cm]{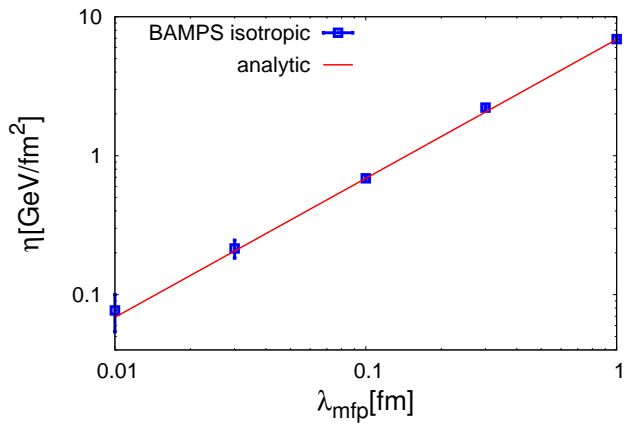}
\caption{Shear viscosity as a function of mean free path.
}
\label{fig:eta}
\end{figure}

In Fig.~\ref{fig:eta} we show the mean shear viscosity 
with the standard deviation as a function of mean free path.
The agreement with the analytical results (line) is perfect.
This confirms the proposed method for extracting the 
shear viscosity for relativistic systems from numerical calculations.

\subsection{pQCD interactions}
\label{sec:pQCD}

In this subsection the results on the shear viscosity are presented for a system 
of gluons. For gluon interactions elastic $gg \to gg$ and inelastic 
$gg \leftrightarrow ggg$ in leading-order pQCD based processes are included.
For a more detailed discussion refere to Ref.~\cite{Xu:2004mz}.

Setups for this case are $L=40$ fm, $v_{wall}=0.5$, and $T=0.4$ GeV.
$L$ has to be chosen appropriately as the mean free path increases with decreasing coupling constant $\alpha_s$.
Running coupling is not implemented in the presented BAMPS calculations.

The extracted mean values of the shear viscosity to entropy ratio $\eta/s$
are given in Table~\ref{table1} and also shown in Fig.~\ref{fig:pQCD}
with the resulting standard deviations of the simulations.
The entropy density is taken by its equilibrium value $s=4n$.
\begin{table}[h]
\begin{center}
\caption{
\label{table1}
$\eta/s$ at various $\alpha_s$.
}
  \begin{tabular}{|c|c|c|c|c|c|c|c|}
  \hline
  $\alpha_s$ 	&0.01&0.03&0.1& 0.2& 0.3& 0.5& 0.6 \\
  \hline
  $\eta /s _{2\rightarrow 2}$	&192.5 $\pm$ 23 &32.6  $\pm$ 3.49 &5.76  $\pm$ 0.63 & 2.25  $\pm$ 0.2 & 1.35   $\pm$ 0.14  & 0.64  $\pm$ 0.064  & 0.55 $\pm$ 0.06 \\
  \hline
  $\eta /s _{2\rightarrow 2, 2\leftrightarrow 3}$	&43.6 $\pm$ 5.2&8.22  $\pm$ 0.66 &0.87  $\pm$ 0.09 & 0.26  $\pm$ 0.03 & 0.17   $\pm$ 0.01	  & 0.1  $\pm$ 0.01  & 0.08 $\pm$ 0.01 \\
  \hline
\end{tabular}
\end{center}
\end{table}
\begin{figure}[h]
\includegraphics[width=8.5cm]{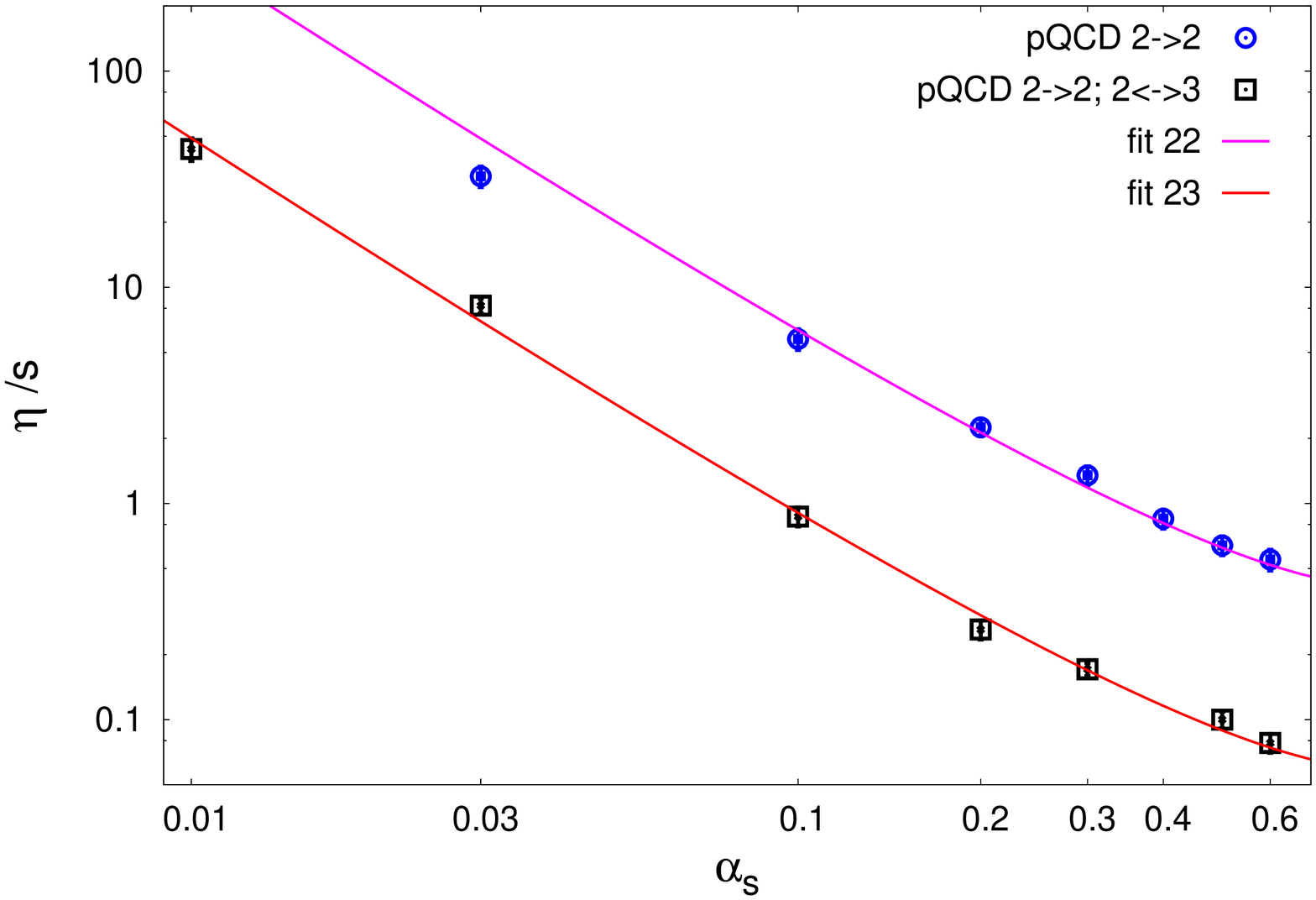}
\caption{Shear viscosity to entropy density ratio at various $\alpha_s$.
}
\label{fig:pQCD}
\end{figure}
\begin{figure}[h]
\includegraphics[width=12cm]{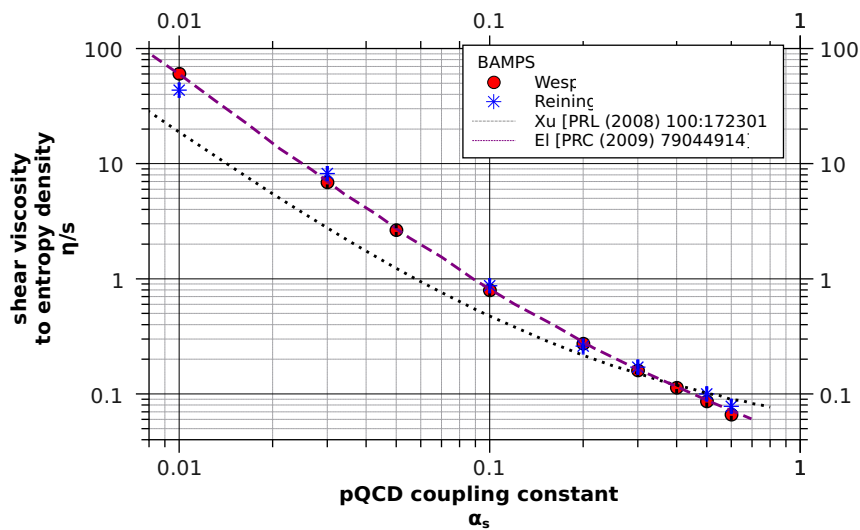}
\caption{Shear viscosity to entropy density ratio at various $\alpha_s$.
Comparisons with other calculations using the same matrix elements
for gluon interactions are made (more in text).
}
\label{fig:pQCD}
\end{figure}

The new results 
are compared to values by Xu et al. \cite{ Xu:2007ns}, El et 
al. \cite{El:2008yy} and Wesp et al. \cite{ Wesp:2008mt}.
Xu et al. identified the shear viscosity coeffient from the 
Navier-Stokes equation and used a gradient expansion in the Boltzmann equation.
Calculating the second moment of the Boltzmann equation they obtained 
the shear viscosity coefficient in terms of the transport collision rate 
defined in Ref.~\cite{ Xu:2007ns}.
El et al. derive the shear viscosity coefficient from the entropy principle, 
which also can be applied to derive the Israel-Stewart equations.
For their derivation El et al. used Grad's approximation for the off-equilibrium 
distribution function and obtained an expression for the shear viscosity similar to the one introduced by Xu et al.,
but with a sligthly different definition of transport collision rate.  
The results from Wesp et al. \cite{Wesp:2008mt} originate from equilibrium fluctuations.
Here the Green-Kubo relations are employed to extract the shear viscosity.
We see a good agreement with all three calculations for $\alpha_s > 0.2$.
For $\alpha_s < 0.2$ our data is in exellent agreement with the data from El
et al. as well es Wesp et al.
We observe the $1/ (\alpha_s^2 log(1/\alpha_s))$ scaling behavior expected from Ref.~\cite{Baym90, AYM}.

Applicability of the methods by Xu et al. and  El et al. crucially 
depends on the chosen parametrisations for the off-equilibrium  
distibution functions.
In particular the momentum dependence of the off-equilibrium correction 
to the equilibrium distribution was chosen differently by these authors, which 
might explain the deviations between their results (Compare also the discussion in Ref.~\cite{Wesp:2008mt}.)

\section{Conclusions and outlook}
\label{sec:conclusions}

In this work we started with the classical picture of shear
viscosity and shear flow. We demonstrated that the classical picture of a
linear velocity field does not apply to relativistic
systems. Rather, we found that velocity fields have a
non-linear form, whereas the rapidity increases in fact linearly.

We also derived an analytical expression for the rapidity
and velocity profiles in systems where the mean-free path
is non-zero. With an increasing mean-free path to system
size ratio the slope of the rapidity profile decreases 
and finite size effects are not negligible anymore.

We employed the numerical transport model BAMPS to create
the velocity and rapidity profiles, compared the numerical results to
our theoretical findings and observed an almost perfect
agreement. The stationary gradient allows us to apply the relativistic
Navier-Stokes equation to calculate the shear viscosity
coefficient $\eta$. We found again a perfect agreement to the analytical
value derived from kinetic theory \cite{deGroot}.
The method proposed here to calculate the shear viscosity coefficient
is thus perfectly suitable for other microscopic transport descriptions.

Furthermore we have then used this setup to calculate 
the shear viscosity to entropy density ratio in
a numerical simulation with elastic and inelastic
pQCD processes implemented in BAMPS for fixed 
coupling constant $\alpha_s$, which is varied from $= 0.01$ to $0.6$.
We compared our results with previously published results 
\cite{ Xu:2007ns, El:2008yy} and also with a very recent work based on the Kubo relation \cite{Wesp:2008mt}
and found a very good agreement.

\section*{Acknowledgements}

The authors are grateful to the Center for the Scientific 
Computing (CSC) at Frankfurt for the computing resources.
FR, CW and IB are grateful 
to ``Helmhotz Graduate School for Heavy Ion research''. 
AE and FR acknowledge support by BMBF.
This work was supported by the Helmholtz International Center
for FAIR within the framework of the LOEWE program 
launched by the State of Hesse.

\appendix{}
\section{}
\label{section_app:wall}
We calculate the number of particles  $\Delta N$ with a thermal Maxwell-Boltzmann distribution
, passing through a wall of area A.
The number of particles passing through a wall orthogonal to the $x$-direction 
in a small timestep $\Delta t$ is equal to the number of all paricles
with distance $\Delta x < - v_x \Delta t$ from the wall:

\begin{equation}
\Delta N = \int_{v_x<0}   \frac{d^3p}{(2 \pi)^3} \int_A dydz \int_{0<x<-v_x\Delta t}dx f(\vec p)=- \int_{v_x<0}\frac{d^3p}{(2 \pi)^3} A v_x \Delta t f(\vec p)
\end{equation}

\begin{equation*}
\frac{\Delta N}{\Delta t} =- \int_{v_x<0}\frac{d^3p}{(2 \pi)^3} A v_x f(\vec p)
\end{equation*}

\begin{equation*}
=-A \int_{v_x<0} \frac{d^3p}{(2 \pi)^3 E} p_x f(\vec p)
=-A \int_{v_x<0} \frac{d^3p}{(2 \pi)^3 E} p_x g e^{-\frac{u^\mu p_\mu}{T}}
\end{equation*}

\begin{equation}
=-\frac{gA}{(2 \pi)^3} \int_{\pi}^{2\pi}d\phi \int_0^{\infty} p_{t} dp_t \int_{-\infty}^{\infty} dy  p_t \cos(\phi) e^{-\frac{p_t \cosh(y+\beta)}{T}}
\end{equation}
where $u^\mu = (\cosh(\beta), 0 , 0, \sinh(\beta) )$. After the  transformation of variables $ y \to y-\beta$ the dependence on the boost velocity drops out:
\begin{equation*}
\frac{\Delta N}{\Delta t}=-\frac{gA}{(2 \pi)^3} \int_{\pi}^{2\pi}d\phi \int_0^{\infty} p_{t} dp_t \int_{-\infty}^{\infty} dy  p_t \cos(\phi) e^{-\frac{p_t \cosh(y)}{T}}
\end{equation*}

\begin{equation*}
=  \frac{2gA}{(2 \pi)^3} \int_0^{\infty}  dp_t \int_{-\infty}^{\infty} dy p_{t}^2 e^{-\frac{p_t \cosh(y)}{T}}
\end{equation*}

\begin{equation*}
=  \frac{2gA}{(2 \pi)^3}  \int_{-\infty}^{\infty} dy  \frac{ 2 T^3}{\cosh^3(y)} =   \frac{gAT^3}{(2 \pi)^2}=\frac{nA}{4}
\end{equation*}
where $n=g T^3/\pi^2$ is the density in the local rest frame.

\newpage


\end{document}